\documentclass[traditabstract,printer]{aa}
\usepackage{txfonts}
\usepackage{graphicx}
\usepackage{color}
\usepackage{natbib}
\usepackage{placeins}
\newcommand\phs{\phantom{$-$}}

\begin{document}
\title{
Unusual neutron-capture nucleosynthesis in a carbon-rich \\Galactic bulge star}
\author{
Andreas Koch\inst{1}  
\and
Moritz Reichert\inst{2}
\and
Camilla Juul Hansen\inst{3,4}
\and 
Melanie Hampel\inst{5} 
\and
Richard J. Stancliffe\inst{6}
\and \\
Amanda Karakas\inst{5}
\and
Almudena Arcones\inst{2,7}
}
\authorrunning{A. Koch et al.}
\titlerunning{Neutron-capture processes in the Galactic bulge}
\offprints{A. Koch; \email{andreas.koch@uni-heidelberg.de}}
\institute{Zentrum f\"ur Astronomie der Universit\"at Heidelberg, Astronomisches Rechen-Institut, M\"onchhofstr. 12, 69120 Heidelberg, Germany
\and Institut f\"ur Kernphysik, Technische Universit\"at Darmstadt, Schlossgartenstr. 2, 64289 Darmstadt, Germany 
\and Max-Planck-Institut f\"ur Astronomie, K\"onigstuhl 17, 69117 Heidelberg, Germany
\and Dark Cosmology Centre, The Niels Bohr Institute, Juliane Maries Vej 30, DK-2100 Copenhagen, Denmark
\and Monash Centre for Astrophysics, School of Phyics and Astronomy, Monash University, 3800, Australia
\and E. A. Milne Centre for Astrophysics, Department of Physics \& Mathematics, University of Hull, HU6 7RX, UK
\and GSI Helmholtzzentrum f\"ur Schwerionenforschung GmbH, Planckstr. 1, 64291 Darmstadt, Germany
}
\date{}
\abstract{Metal-poor stars in the Galactic halo often show strong enhancements in carbon and/or neutron-capture elements. 
However, the Galactic bulge is notable for its paucity {of these} carbon-enhanced metal-poor (CEMP) and/or CH-stars, with only two such objects known to date. This begs the question whether the 
processes that produced  their abundance distribution 
were governed by a comparable nucleosynthesis in similar stellar sites as for their more numerous counterparts in the halo. 
Recently, two contenders of these classes of stars were discovered in the bulge,  
at [Fe/H] = $-1.5$ and $-$2.5 dex, both of 
which show enhancements in [C/Fe] of 0.4 and 1.4 dex {(respectively)},  [Ba/Fe] in excess of 1.3 dex, and also elevated {nitrogen}. 
The more metal-poor of the stars can be well matched by standard $s$-process nucleosynthesis in low-mass 
Asymptotic Giant Branch (AGB) polluters. The other star shows an abnormally high [Rb/Fe] ratio.
Here, we further investigate the origin of the abundance peculiarities in the Rb-rich star by new, detailed measurements of heavy element abundances and by 
comparing the chemical element ratios of 36 species to 
 several models of neutron-capture nucleosynthesis. 
The $i$-process with intermediate neutron densities between those of the slow  ($s$-) and rapid ($r$)-neutron-capture processes has been previously found to provide
good matches of CEMP stars with enhancements in both $r$- and $s$-process elements (class CEMP-$r/s$), rather than invoking a superposition of yields from the 
respective individual processes. However, the peculiar bulge star is incompatible with a pure $i$-process from a single ingestion event. 
Instead, it can, { statistically, } be better reproduced by { more convoluted} models accounting for two proton ingestion events, or by an $i$-process component in combination with
$s$-process nucleosynthesis in low-to-intermediate mass (2--3 M$_{\odot}$) AGB stars, indicating multiple polluters.
Finally, we discuss the impact of mixing during stellar evolution on the observed abundance peculiarities. 
}
\maketitle
%
%
%
%
\section{Introduction}
%
Nuclei heavier than Z$\gtrsim$30 can be created  via neutron-captures, which 
can be separated into the slow ($s$) and rapid ($r$) neutron-capture processes  \citep{Burbidge1957}, 
as determined by the relative efficiency of the capture rates versus competing beta-decay timescales. 
{ {Rare isotopes of heavy elements} are formed by neutron captures with large cross sections, 
or by disintegration reactions or  various other flavours of the p-process such as ($p,\gamma$) reactions.}
Since the slow and rapid processes require significantly different neutron densities, they have accordingly been assigned to different sites such as asymptotic giant branch (AGB) stars 
for the $s$-process \citep[e.g.,][]{Busso1999,Gallino1998,Kaeppeler2011,Karakas2014} vs. supernova (SN) nucleosynthesis \citep{QianWasserburg2007,Sneden2008,Winteler2012} or neutron star mergers for the $r$-process \citep[e.g.,][]{Lattimer1974,Freiburghaus1999,Chornock2017,Watson2018}.

Recent observations suggested the need for an additional process acting at conditions  between $s$ and $r$, viz. the intermediate neutron-capture process ($i$-process), originally proposed by \citet{CowanRose1977}.
Rather than invoking a pollution of the interstellar material with both $r$- and $s$-enhanced material from different sites to account for the abundance peculiarities seen in, e.g., a sub-class of carbon-enhanced 
metal-poor stars (CEMP $r/s$\footnote{That is, CEMP stars with strong enhancements in $r$- and $s$-process elements \citep{BeersChristlieb2005}.}), 
the $i$-process acts at neutron densities between the $r$- and $s$-process in a single site, thus producing a markedly different abundance 
pattern\footnote{It is not straightforward to  draw a  
distinction between the $i$- and $s$-processes at one definitive value for the neutron density. 
\citet{Fishlock2014} suggested that, while  $N$ exceeded 10$^{13}$  cm$^{-3}$ in their intermediate-mass AGB models, 
 the resulting abundance distribution was still very much that of  an $s$-process, while being
 dominated by first peak elements due to \mbox{$^{22}$Ne($\alpha$, $n$)$^{25}$Mg}  providing the neutrons. 
The reality is more likely that there is some overlap between neutron densities that are classically considered 
"$s$" and are "$i$". {Moreover, a distinction in terms of the $\tau$--$n$-density space occupied by these two processes may be possible.}}. 
{Calculations of the $i$-process}  are able to provide good fits to observations of strongly enhanced Ba and Eu abundances and in particular to reproduce the stars' 
high [$hs$/$ls$] ratios\footnote{The ratio of heavier, second-peak $s$-process elements to the lighter, first-peak elements. In the following we will adopt 
[$ls$/Fe]= {[Sr+Y+Zr/Fe]/3 and  [$hs$/Fe]=[Ba+La+Ce/Fe]/3}; e.g., \citet{Cristallo2011,Abate2015stat}.
We further follow the usual spectroscopic notation in terms of the number densities 
$N_A$ and $N_B$ for elements A and B, relative to the Sun: [A/B] = $\log_{10}(N_A/N_B)\,-\,\log_{10}(N_A/N_B)_{\odot}$.} \citep[e.g.,][]{Hampel2016,Denissenkov2018}. 
Indications of $i$-process signatures have been observed in 
grains \citep{Jadhav2013}, 
{post-AGB stars} \citep{Lugaro2015}, 
open-cluster stars \citep{Mishenina2015}, 
CEMP stars \citep[e.g.,][]{Hampel2016}, 
the most metal-poor stars known \citep{Clarkson2018}, 
and  a carbon-normal, 
metal-poor field dwarf with enhanced $s$- and $r$-process abundances \citep{Roederer2016iproc}. 
Proposed sites for $i$-process nucleosynthesis are, amongst others, the {He-core and He-shell flashes in low-mass, low-metallicity stars} \citep{Campbell2008,Campbell2010,Cristallo2009,Stancliffe2011}, 
Super-AGB stars \citep{Doherty2015,Jones2016iproc}, and rapidly accreting white dwarfs \citep{Denissenkov2017}. 

Here, we investigate the nucleosynthetic signatures of a metal-poor ([Fe/H]=$-1.5$ dex) star 
in the Galactic bulge that shows evidence of strong $s$-process enhancements \citep{Koch2016} without indication for strong over-abundances of the $r$-process elements.
This CH-star shows a peculiar signature of two abundance peaks with similar enhancements, namely around Rb (Z=37)\footnote{Extremely Rb-rich, self-enriched 
AGB stars have been reported to exist \citep{GarciaHernandez2006,Zamora2014}, but no detailed abundance distributions are available for those objects.}
 and Ba (Z=56). 
In \citet{Koch2016} we found that 
the abundance pattern of this star {suggested} enrichment from an intermediate mass AGB star of $\sim$4 M$_{\odot}$,  although
the entire distribution could not be satisfactorily fitted. Such a deficiency of standard $s$-process nucleosynthesis prompts the need for further 
complexity in the form of admixing other nucleosynthetic channels.  
{We therefore perform} a detailed comparison of the observed abundance pattern in this bulge CH-star with calculations of $s$- and $r$-processes, combined with predictions from $i$-process nucleosynthesis. 

This paper is organized as follows. In Sect.~2 we place this object in the context of other C-rich stars in the Milky Way's components 
and we recapitulate the observed abundance details that are complemented with new measurements of several heavy elements; 
in Sect.~3 we  introduce the   $s$, $r$, and $i$-process models used to represent the targets' abundance patterns, while, in Sect.~4, we describe the best-fit models to investigate, which processes 
dominated this bulge star's enrichment.
To improve the results, we consider enrichment from multiple sites in Sect.~5 and discuss alternative scenarios in Sect.~6. 
 Finally, Sect.~7 summarizes our findings. 
\section{Metal-poor bulge stars}
While the Galactic bulge is predominantly old and metal-rich \citep{McWilliamRich1994,Clarkson2008,McWilliam2016}, recent studies have focused on the detection and analysis of metal-poor
stars towards the bulge, which are predicted by cosmological models to reside in those central regions \citep[e.g.,][]{Tumlinson2010}. 
In fact, \citet{CaseySchlaufman2015} measured depleted [Sc/Fe] ratios in three metal-poor bulge stars, which they interpreted as a signature of enrichment by the first, massive Population III stars, 
while no other such sample shows any such evidence \citep{Koch2016}. 
\subsection{Carbon-rich bulge stars}
In \citet{Koch2016} we detected two stars with strong carbon enhancements towards the Galactic bulge. Subsequent analyses classified them as 
a CEMP-$s$ star (star-ID \#27793\footnote{Following the naming scheme of \citet{Koch2016}. The IAU names for these objects are
J183113.29-335148.3 (=\#27793) and J183003.87-333423.6 (=\#10464).}; 
[Fe/H]=$-2.52$; [C/Fe]=1.44; [Ba/Fe]=1.31)
and a moderately metal-poor CH-star (star-ID \#10464; [Fe/H]=$-1.53$; [C/Fe]=0.41; [Ba/Fe]=1.35). 
These are the first known contenders of these classes of stars in the Galactic bulge. 

In order to understand the origin of these stars' abundance pattern and to connect it to any peculiar class of objects
it is indispensable to detect and characterize more, similar candidates. 
However,  so far no other CH- or CEMP-stars have been found in the bulge,  
save for very few of their metal-rich counterparts, the Ba-stars \citep{Lebzelter2013}, 
that follow the dominant metallicity distribution function (MDF) of the bulge. 
The target of the present study (\#10464) and the bulge CEMP-$s$ star \#27793 \citep{Koch2016} are such rare exceptions. 
This keeps the fraction of CEMP stars in the bulge down at the 2\% level. 
The reason for this can be sought in the currently observed, 
overall, more metal-rich nature of the bulge, albeit theories predict the occurrence of such very metal-poor stars towards the Galactic
centre regions \citep[e.g.,][]{Tumlinson2010,Ness2013,CaseySchlaufman2015,Koch2016}. 

The fraction of CEMP stars in the halo and in metal-poor dwarf spheroidal galaxies is known to significantly increase
with decreasing metallicity \citep[e.g.,][]{Norris2010Seg1,Carollo2012,Salvadori2015,Skuladottir2015,TTHansen2015,CJHansen2016,Susmitha2017}
and also the bulge's 
metal-poor population can be expected to follow this trend \citep{Tumlinson2007,Tumlinson2010}. 
It is then interesting to note that the bulge CH-star falls right on the peak of the halo MDF, and the CEMP-$s$ lies at the peak of the halo CEMP star distribution.
Moreover, the commonly accepted scenario for the origin of the $s$-process enhancements in the CEMP-$s$ stars is mass transfer from 
an (AGB) companion in a binary system \citep[e.g.,][]{Bisterzo2011,Starkenburg2014,TTHansen2016}. 
Apart from the obvious contenders such as survey target selection biases \citep{Jacobson2015}, evolutionary mixing on the red giant branch that depletes
the surface abundance of C \citep{Placco2014}, and overall low number statistics of metal-poor bulge stars \citep[e.g.,][]{Koch2016}, 
the present paucity of bulge CEMP-$s$ stars could therefore also bear implications for the  binary fraction in the early bulge, which to date has been difficult 
to determine \citep[e.g.,][]{Holtzman1998,Miszalski2009}. 
 On the other hand, surveys to date failed to detect even the  CEMP-no stars  in the bulge \citep[e.g.,][]{Howes2016}; in the Galactic halo, 
this subclass, not over-enhanced in any of the heavy elements\footnote{Sr may, relatively speaking, have  higher abundance ratios than, e.g., Ba, 
but typically it still shows abundance ratios that are at most mildly elevated to [Sr/Fe]$\lesssim$0.5 \citep{Yong2013}, but primarily Solar or below \citep{CJHansen2016, CJHansen2019} 
in CEMP-no stars.}, are not part of binary systems \citep{CJHansen2016}
so the very low fraction of CEMP stars of any class in the bulge indicates that their absence has {multiple origins rather} than only being related to the bulge binary fraction, 
which can differ from that of the halo \citep[e.g.,][]{Ryan1992}. 
\subsection{Chemical peculiarities in metal-poor bulge stars}
In \citet{Koch2016}, comparison with standard AGB yields \citep{Cristallo2011} indicated that the C- and $s$-process enhancements in the regular CEMP-$s$ star \#27793 were best matched with 
mass transfer from a low-mass AGB companion, although details of the AGB nucleosynthesis such as the size of the $^{13}$C-pocket and mass loss suggest a more complicated picture. 
More complications arose in the attempt to reproduce the abundance pattern of the target of this present study, star \#10464, which shows contributions from AGB nucleosynthesis. 
However, no satisfactory fit to the $hs$- and simultaneously the $ls$-peak elements could be obtained, leaving a large uncertainty beyond the ``low-to-intermediate mass AGB'' 
enrichment. Here, the largest deviation from model fits \citep[e.g.,][]{Cristallo2011} occurred for Rb, which, at [Rb/Fe] = 1.29$\pm$0.16 dex, remains inexplicably high. 
\subsection{Stellar parameters and additional abundance measurements}
The stellar parameters  of the peculiar object \#10464 we found in \citet{Koch2016} are 
(T$_{\rm eff}$, log\,$g$, $\xi$, [Fe/H]) = (5400 K, 1.7, 2.64 km\,s$^{-1}$, $-1.53$). 
{ 
In that work, as well as in the following, we had performed an abundance analysis using the  LTE abundance code MOOG \citep{Sneden1973} and building on the 
plane-parallel, one-dimensional grid of ATLAS model atmospheres\footnote{\url{http://kurucz.harvard.edu/grids.html}}.
This choice is adequate for star \#10464 as it is a non-variable Horizontal Branch star 
and furthermore, a proper modeling of dynamic atmospheres is non-trivial and thus, to date, often approximated by static theory 
(\citealt{Hansen2016RR,Vasilyev2018}; cf. \citealt{GarciaHernandez2007}). We also note that our working hypothesis is that 
the unusual chemical abundances found in the present-day star are the product of nucleosynthesis in a long-perished generation of  polluters. The 
evolutionary state of the latter, whether with strong atmospheric dynamics or not,  is thus irrelevant for the abundance derivation in the present object. 

In \citet{Koch2016}, we employed an equivalent width analysis and enforced excitation and ionization balances to obtain the stellar parameters. 
Here, we verified these parameters using the novel code ATHOS (``A Tool for HOmogenizing Stellar parameters'', \citealt{Hanke2018}), which 
uses flux ratios within an optimized set of spectral ranges. 
The resulting temperature and metallicity are in excellent agreement with the previous results. 
The gravity returned by ATHOS is marginally lower, but as Table~7 of \citet{Koch2016} indicates, this has only a minor influence on the derived
abundances ratios. In particular, as a neutral species, Rb is highly insensitive to this parameter. Therefore we conclude that the set of stellar parameters 
we use in this work is reliable. 
}

Table~1 recapitulates the abundance measurements in this star obtained in the latter work.  
In addition, we were able to extract further elemental abundances not included in the latter study. 
Here, we also list the total error bar on our measurements, which is based on 
the contribution from the statistical and systematic uncertainties. The former was based on the 1$\sigma$-scatter of lines in case that several transitions were
measurable, and estimated from the quality of the fitting procedures otherwise. Systematic errors, in turn, were derived from the standard technique of 
varying the stellar models by one parameter about its uncertainty at a time, thereby re-deriving a new set of abundances \citep{Koch2016}.  
This full set of abundances will be the basis of our comparison with various models in Sect.~4. 
\begin{table*}[!t]
\caption{Abundance ratios in the C-rich bulge star \#10464 from \citet{Koch2016} and our present measurements.}
\centering
\begin{tabular}{lcc|lcc|lcc|lcc}
\hline\hline	   
Element & log\,$\varepsilon$ & [X/Fe]\tablefootmark{a}  &
Element & log\,$\varepsilon$ &  [X/Fe]\tablefootmark{a}  &
Element & log\,$\varepsilon$ &  [X/Fe]\tablefootmark{a}  &
Element & log\,$\varepsilon$ &  [X/Fe]\tablefootmark{a}  \\
\hline  		
Li\,{\sc i}  & & $\llap{$<$}$0.70$\phantom{\pm0.00}$    & Cr\,{\sc i}  & 3.83 & $\llap{$-$}0.28\pm0.07$  & Zr\,{\sc ii} & 	     1.35  & $0.30\pm0.17$ & Ho\,{\sc ii} & $\llap{$-$}0.65$ &           $0.40\pm0.20$ \\        
 C\,{\sc i}  &   7.31 &      $0.41\pm0.21$		& Mn\,{\sc i}  & 3.64 & $\llap{$-$}0.26\pm0.14$  & Ba\,{\sc ii} & 	     2.00  & $1.35\pm0.08$ & Er\,{\sc ii} & $\llap{$-$}0.01$ &           $0.60\pm0.20$ \\  
 N\,{\sc i}  &   7.05 &      $0.75\pm0.15$		& Fe\,{\sc i}  & 5.97 & $\llap{$-$}1.53\pm0.06$  & La\,{\sc ii} & 	     0.49  & $0.92\pm0.15$ & Hf\,{\sc ii} &            0.82  &           $1.50\pm0.20$ \\  
 O\,{\sc i}  &   7.79 &      $0.63\pm0.13$		& Fe\,{\sc ii} & 5.98 & $\llap{$-$}1.52\pm0.06$  & Ce\,{\sc ii} & 	     1.30  & $1.24\pm0.17$ & Pb\,{\sc i}  &	       1.72  &           $1.50\pm0.20$ \\
\cline{10-12}
Na\,{\sc i}  &   5.25 &      $0.54\pm0.06$		& Co\,{\sc i}  & 3.50 &           $0.04\pm0.09$  & Pr\,{\sc ii} &            0.11  & $0.92\pm0.05$ & [C/N]        & 	$\ldots$      & $\llap{$-$}0.34\pm0.26$ \\
Mg\,{\sc i}  &   6.56 &      $0.49\pm0.08$		& Ni\,{\sc i}  & 4.73 &           $0.03\pm0.10$  & Nd\,{\sc ii} &            1.02  & $1.13\pm0.10$ & [N/O]        & 	$\ldots$      & 	  $0.12\pm0.20$ \\
\cline{10-12}
Si\,{\sc i}  &   6.56 &      $0.58\pm0.07$		& Zn\,{\sc i}  & 3.34 &           $0.31\pm0.05$  & Sm\,{\sc ii} &            0.33  & $0.90\pm0.09$ & [Ba/La]      & 	$\ldots$      & 	  $0.43\pm0.17$ \\					  
Ca\,{\sc i}  &   5.00 &      $0.19\pm0.10$		& Ga\,{\sc i}  & 2.52 &           $0.40\pm0.20$  & Eu\,{\sc ii} & $\llap{$-$}0.64$ & $0.37\pm0.16$ & [Eu/La]      & 	$\ldots$      & $\llap{$-$}0.55\pm0.22$ \\					 
Sc\,{\sc ii} &   1.62 &      $0.00\pm0.08$		& Rb\,{\sc i}  & 2.28 &           $1.29\pm0.16$  & Gd\,{\sc ii} &            0.09  & $0.55\pm0.11$ & [$hs$/Fe]    & 	$\ldots$      & 	  $1.17\pm0.08$ \\  
Ti\,{\sc i}  &   3.79 &      $0.37\pm0.09$		& Sr\,{\sc ii} & 2.18 &           $0.84\pm0.07$  & Tb\,{\sc ii} & $\llap{$-$}0.38$ & $0.85\pm0.20$ & [$ls$/Fe]    & 	$\ldots$      & 	  $0.53\pm0.07$ \\
 V\,{\sc i}  &   2.22 & $\llap{$-$}0.18\pm0.16$		&  Y\,{\sc ii} & 1.14 &           $0.46\pm0.11$  & Dy\,{\sc ii} & $\llap{$-$}0.05$ & $0.38\pm0.10$ & [$hs$/$ls$]  & 	$\ldots$      & 	  $0.64\pm0.11$ \\
\hline
\end{tabular}
\tablefoot{\tablefoottext{a}{The given, total error includes a 1$\sigma$ statistical and the systematic uncertainty.}}
\end{table*}

In our previous work, the C-abundance of this star had been derived by spectral synthesis of the CH G-band at 4300 \AA, yielding a [C/Fe] ratio of 0.4 dex. 
Here, from spectral synthesis of the CN-band at 3883 \AA, we derived a nitrogen abundance ratio of [N/Fe]=0.75$\pm$0.15, with an uncertainty mainly driven by the continuum placement. 
The low [C/N] of $-0.34\pm0.19$ dex in this CH-star is close to the 
limit that separates mixed and unmixed metal-poor stars
\citep{Spite2005,CJHansen2016}. This will be further discussed in Sect.~6.
In spite of its larger [N/Fe] ratio in excess of 0.5 dex, the [C/N] ratio of star \#10464 is marginally too high for it to qualify as a ``Nitrogen-enhanced metal-poor'' 
 star
\citep{Johnson2007NEMP,Pols2012}.
An O-abundance from the triplet lines at 7770 \AA~could be determined and yielded a value of [O/Fe]=0.63$\pm$0.13 dex.

The blue spectral range of our spectra allowed us to complement our earlier, basic abundance ratios by a wealth of 
measurements for neutron-capture elements \citep{CJHansen2015}. To this end, we employed spectral synthesis for stronger lines 
that were chosen from the list of \citet{Roederer2014}. Hyperfine structure was included where appropriate, and  a line list providing 
the base for the additional measurements is given in Table~2. 
{ Thus we were able to determine additional abundances for Li, N, O, Ga, Ce, Pr, Sm, Gd, Tb, Dy, Ho, Er, Hf, and Pb that were not 
included in our original work \citep{Koch2016}.
}
\begin{table*}[htb]
\caption{Line list for heavy elements in \#10464 that were not covered in \citet{Koch2016}.}
\centering          
\begin{tabular}{cccr|cccr|cccr}
\hline\hline       
Element & {$\lambda$ [\AA] } &  {E.P. [eV]} &  {log\,$gf$} &
Element & {$\lambda$ [\AA] } &  {E.P. [eV]} &  {log\,$gf$} &
Element & {$\lambda$ [\AA] } &  {E.P. [eV]} &  {log\,$gf$}  \\
\hline
Li\,{\sc I}  & 6707.80 & 0.00 & \phs0.17 & Sm\,{\sc II} & 4815.81 & 0.19 &  $-$0.77 & Dy\,{\sc II} & 3757.37 & 0.10 &  $-$0.17 \\
 O\,{\sc I}  & 7771.94 & 9.15 & \phs0.32 & Pr\,{\sc II} & 4062.80 & 0.42 & \phs0.33 & Dy\,{\sc II} & 3944.68 & 0.00 & \phs0.11 \\
 O\,{\sc I}  & 7774.17 & 9.15 & \phs0.17 & Pr\,{\sc II} & 4141.22 & 0.55 & \phs0.38 & Dy\,{\sc II} & 4103.31 & 0.10 &  $-$0.38 \\
 O\,{\sc I}  & 7775.39 & 9.15 &  $-$0.05 & Pr\,{\sc II} & 4143.13 & 0.37 & \phs0.60 & Dy\,{\sc II} & 4449.70 & 0.00 &  $-$1.03  \\
Ga\,{\sc I}  & 4172.00 & 0.10 &  $-$0.31 & Pr\,{\sc II} & 4179.40 & 0.20 & \phs0.46 & Ho\,{\sc II} & 3810.71 & 0.00 & \phs0.19  \\
Ce\,{\sc II} & 5274.23 & 1.04 & \phs0.15 & Pr\,{\sc II} & 4222.95 & 0.06 & \phs0.23 & Ho\,{\sc II} & 4045.45 & 0.00 &  $-$0.05 \\
Sm\,{\sc II} & 4536.51 & 0.10 &  $-$1.28 & Pr\,{\sc II} & 4408.81 & 0.00 & \phs0.05 & Er\,{\sc II} & 3692.65 & 0.05 & \phs0.14  \\
Sm\,{\sc II} & 4577.69 & 0.25 &  $-$0.65 & Gd\,{\sc II} & 4130.37 & 0.73 &  $-$0.02 & Er\,{\sc II} & 3729.52 & 0.00 &  $-$0.59  \\
Sm\,{\sc II} & 4642.23 & 0.38 &  $-$0.46 & Gd\,{\sc II} & 4251.57 & 0.38 &  $-$0.22 & Hf\,{\sc II} & 4093.16 & 0.45 &  $-$1.15  \\
Sm\,{\sc II} & 4676.90 & 0.04 &  $-$0.87 & Tb\,{\sc II} & 4752.53 & 0.00 &  $-$0.55 & Pb\,{\sc I}  & 4057.81 & 1.22 &  $-$0.22  \\
\hline
\hline
 \end{tabular}
\end{table*}

To illustrate the range of our measurements, we show in Fig.~1 the full abundance pattern for \#10464 together with { an exemplary} range of AGB models from the F.R.U.I.T.Y. database \citep{Cristallo2011}, highlighting the difficulty in simultaneously reproducing all heavy-element peaks, in particular the star's high [Rb/Fe] ratio. 
\begin{figure}
\begin{center}
\includegraphics[angle=0,width=1\hsize]{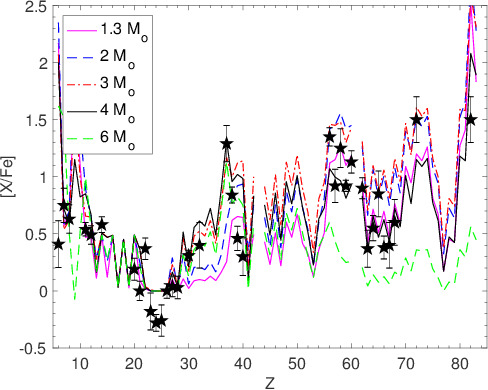}
\includegraphics[angle=0,width=1\hsize]{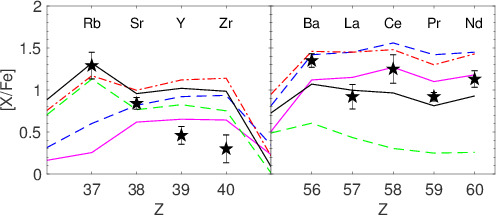}
\end{center}
\caption{Full abundance distribution of the bulge star \#10464 from \citet{Koch2016} and this work. The AGB yields with progenitor masses as labeled were taken from \citet{Cristallo2011}.
The bottom panels are a zoom into the regions of the first and second $n$-capture element peaks.}
\end{figure}
\section{Model details of $s$-,  $r$-, and  $i$-process nucleosynthesis}
\subsection{$s$-process yields from AGB nucleosynthesis}
Here, we employed the metal-poor (${Z}$ = 0.0001; [M/H]=$-2.2$ dex) AGB models of 
\citet{Lugaro2012}. 
{In order to determine the best model to describe this star, we applied our fitting routines (Sect.~4) 
to the entire, broad range of stellar masses (0.9--6 M$_{\odot}$) provided by these models.} 
This set of AGB calculations also accounted for varying initial chemical compositions 
{(e.g., in terms of varying heavy element contributions from early Galactic chemical enrichment, \citealt{Kobayashi2006})} 
and one of the,  
to date, still most uncertain parameters in AGB nucleosynthesis -- the 
size of the $^{13}$C pocket {(see, e.g., \citealt{Buntain2017} for a detailed discussion)}.
Observations indicate a variety of pocket sizes; the convective boundary mixing that is relevant for transporting 
H into the intershell is not well understood, even though many processes have been proposed.

Similar tests were carried out using the more metal-rich models of 
\citet[][${Z}$=0.001]{Fishlock2014} and \citet[][${Z}$=0.007 and ${Z}$=0.014]{KarakasLugaro2016}, but those resulted in 
considerably larger $\chi^2$ values when fitted to the observations. Coupled with the low metallicity of the star to be described, at [Fe/H]=$-1.5$ dex, 
we discard these metal-rich yields from the following considerations. 
{ The overall, best-fit $s$-process-alone model (viz. a 3~M$_{\odot}$ AGB) will be described in detail in Sect.~4.1.}
\subsection{$r$-process yields from neutron star mergers}
{The r-process calculation} was performed with the WinNet nucleosynthesis network \citep{Thielemann2011,Winteler2014} that contains almost $6000$ nuclei, using $\sim 65000$ reaction rates of the Jina Reaclib Database V2.0 \citep{Cyburt2010}, using the Finite-Range droplet mass model \citep{Moeller1995}. In addition, we used neutron capture and neutron-induced fission rates given by \citet{Panov2010}. For temperatures lower than $T \le 0.01\, \mathrm{GK}$ some electron-capture and $\beta$-decay rates are replaced by the ones of \citet{Langanke2001}. Here, the $r$-process is calculated in the environment of dynamical ejecta from compact neutron star mergers using temperature and density profiles from  the 
Newtonian simulations of \citet{Price2006}.  \citet{Korobkin2012} investigated the $r$-process nucleosynthesis for these ejecta and found a very robust abundance pattern for heavy nuclei, caused by the low electron fractions of $\sim 0.035$ that leads to fission cycling. Therefore, we choose one representative temperature and density profile to calculate the typical $r$-process abundances. Despite recent direct evidence for $r$-process nucleosynthesis in neutron star mergers \citep[e.g.,][]{Chornock2017,Watson2018} 
other hosts are also promising astrophysical sites, e.g. magnetohydrodynamically driven supernovae. {Besides uncertain astrophysical conditions, most of the nuclear reactions involved in the $r$-process nowadays still rely on theoretical predictions rather than experimental data. As a consequence, theoretical nucleosynthesis predictions are not able to fully reproduce the solar $r$-process abundances. Therefore, we also considered the abundance pattern of the metal-poor star CS 22892-052 as a reference set, assuming that its heavy elements are produced by the $r$-process only \citep{Sneden2003}. Even with this pattern we reach the same conclusion that the contribution of $r$-process to the bulge star \#10464 is negligible, as also illustrated below in Fig.~4.}

\subsection{Basics of the $i$-process}
The $i$-process is thought to occur when H is advected into a convective zone that is driven by helium burning. Hydrogen reacts with $^{12}$C to produce $^{13}$N, which can decay to 
$^{13}$C. The latter isotope finally reacts with the available He via $^{13}$C($\alpha$,n)$^{16}$O reaction, producing the necessary neutrons. Contrary to $^{13}$C-pockets in regular AGB models, which hold all material including heavy ($s$-process) elements in very localized regions, freshly produced nuclei in common $i$-process models can be distributed throughout the entire convective zone. The high temperatures lead to rapid neutron production and a characteristic
neutron density of $\sim$$10^{14}$--$10^{15}$ cm$^{-3}$
\citep{CowanRose1977,Herwig2011,Roederer2016iproc,Hampel2016}. 
The energy release through these hydrogen-burning chains 
 could lead to the expelling of the envelope and self-quenching \citep{Jones2016iproc}. 
This terminates the $i$-process, although the  time until termination will likely vary between different stellar sites, down to  the order of days as in the case of Sakurai's object \citep{Herwig2011}.

{ There is as yet no concrete site for the i-processes and many have been proposed, including proton ingestions in AGB stars, 
very late thermal pulses, Super-AGB stars and rapidly 
accreting white dwarfs. While we are not attempting to choose between them here,  the physical conditions in our models 
were based on the first the ones mentioned above. 
Similarly, the metallicity constraints of this process are not clear, yet: observationally, the $i$-process is seen to operate at very low metallicity
(CEMP-$i$ stars; \citealt{Hampel2016}), but there is also evidence at higher metallicities, e.g., in Sakurai's object that shows in-situ 
$i$-process nucleosynthesis at solar metallicity \citep{Herwig2011}. The latter situation is different from the star of our present analysis, which 
has not produced its heavy elements itself, but it had formed out of material enriched in these metals by a previous generation of events.
}
\subsection{Equilibrium $i$-process with fixed neutron exposure}
Using the suite of {codes {\em NucNet Tools} } \citep{Meyer2012NIC} we created a one-zone model 
with given initial composition under conditions of fixed temperature and density. The latter were chosen as representing the mid-point of the 
intershell region in a low-metallicity AGB model (see \citealt{Stancliffe2011} for further details of the structure), at values of 
$T = 1.5 \times 10^8$ K and $\rho = 1600$ g\,cm$^{-3}$. The initial chemical composition of this region 
represents the intershell of a low-metallicity (${Z}$ = 0.0001), low-mass (M = 1 M$_{\odot}$) AGB model after the second thermal pulse
\citep[][and references therein]{Abate2015models}.
The nuclear network was  followed with 5442 isotopes and 45831 reactions from the JINA Reaclib V0.5 database \citep{Cyburt2010} 
with further  {$\alpha$-}decay rates from Tuli (2011)\footnote{{http://www.nndc.bnl.gov/wallet/wc8.html}}.

The evolution of the abundance distribution was followed 
at a fixed neutron density  of $N=10^{15}$ cm$^{-3}$ 
for 0.1 years,  
which resulted in a neutron exposure of $\tau = 495$ mbarn$^{-1}$. 
This short time was sufficient to result in an equilibrium abundance pattern of the heavy elements, independently of the initial chemical 
composition. 
 As elaborated in \citet{Hampel2016}, the runtimes of models at lower neutron densities (down to 10$^{-9}$ cm$^{-3}$) were scaled with $N$ to 
ensure the same neutron exposure.
While this chosen exposure has the advantage of ensuring the robust equilibrium-abundance pattern, it has the drawback that Pb (Z=82) abundances cannot be predicted, because the reaction flows around lead cannot reach equilibrium 
 -- this results in an implausibly high level of Pb production despite the otherwise very robust $i$-process pattern.
  Details of the shortcomings in a proper prediction of Pb are discussed in detail in \citet{Hampel2016}.
While lead represents, alone, the third $s$-process peak and thus is useful to verify the 
robustness of nucleosynthesis models, 
we explicitly removed Pb from all further consideration in our statistics given the above complications with its modeling.

The run-times of all other models with differing neutron densities 
from $N=10^{9}$ cm$^{-3}$ to $N=10^{15}$ cm$^{-3}$  
were finally  scaled with neutron density to ascertain the same, constant neutron exposure in all models. 
to allow for comparisons of the different equilibrium patterns. 
We refer the reader to \citet{Hampel2016} for details on the $i$-process models.  
\section{Model results}
Here, we apply the setups laid out in Sect.~3 to test if the bulge star \#10464  shows signatures indicative of
 $s$-, $r$-, or $i$-process nucleosynthesis, or combinations thereof.
{ We assume that the nucleosynthetic processes described did not take place in this star itself, but occurred in an earlier generation 
which polluted the gas from which this star formed.}
In the following, we only consider elements with $32 \le$ Z $\le 72$ 
{ (Ga through Hf} in the statistics, since nuclei below Z$\la30$ 
are not significantly produced in the neutron-capture reactions
and in order to properly model  light isotopes (such as $^{13}$C or $^{14}$N), a careful treatment of the entrainment and nucleosynthesis processes 
in multi-dimensional simulations is needed  \citep{Herwig2011,Herwig2014}. 
We note, however, that the C-, \mbox{N-,} and O-abundances in this star agree very well to within the uncertainties with model predictions of, e.g., 
the AGB models detailed in Sect.~3.1. 
{ As explicated above,  Pb was excluded from our statistics as well.}

The results of the fitting are summarized in Table~3, where the mass of the $s$-process contributing AGB-star is indicated in Solar masses as a subscripts (ala ``$s_{m2}$'' for a 2 M$_{\odot}$ star
and the $i$-process is identified by the log of its neutron density. 
The quality of each scenario was judged in terms of the $\chi^2$ statistics for each of the enrichment scenarios.
This statistical estimator, within the element range of $Z_i \le  Z \le Z_f$,  is given by
\begin{equation}
\chi^2 =  \sum_{Z_i}^{Z_f} \left( \, \log\left( \varepsilon(Z) \right) - \log\left( c\cdot Y \right)  \, \right) ^2 / \, \sigma(Z)^2
\end{equation}
where $\sigma(Z)$ is the error on the observationally derived  $\log \varepsilon$  abundances (Table~1),  
and $Y$ are the model yields from either process.
The fit of one distribution is obtained by a multiplicative scaling factor $c$ of the abundances 
$Y$, which translates into an additive scaling in logarithmic space. This is equivalent to an admixture of the individual processes with pure 
hydrogen. 

{ Under the assumption of Gaussian errors and considering that we have N=18 elements in our fit range of 32$\le$Z$\le$72, we
can estimate that a statistically good fit corresponds to a $\chi^2$ of about 40, while an excellent result should yield values on the order of 10.} 

The resulting abundance distributions
for a chosen set are shown in Figs. 2--4. 
As Table~3 implies, admixtures of the Solar abundance distribution \citep{Lodders2003} has an adverse effect on the statistics and we do not
consider this option any further. 
\begin{table}[!btp]
\caption{Results for various linear combinations of nucleosynthetic processes in the fitting range of \mbox{$32 \le \mathrm{Z}\le 72$.}}
\centering
\begin{tabular}{lcc|lcc}
\hline\hline       
{Process(es)}  &  $\chi^2$ &  Notes    & {Process(es)}  &  $\chi^2$ &  Notes  \\
\hline
$\mathrm{solar}$                         & 154.72 & 1         &   $s_{m2}+i_{\mathrm{n9}}$		    & 66.31   & 3,2\\		
$i_{\mathrm{n9}}$                        & 214.25 & 2         &   $s_{m2}+i_{\mathrm{n15}}$		    & 51.17   & 3,2\\
$i_{\mathrm{n10}}$                       & 229.76 & 2         &   $s_{m5}+i_{\mathrm{n9}}$		    & 59.25   & 5,2\\	
$i_{\mathrm{n11}}$                       & 251.94 & 2         &   $s_{m5}+i_{\mathrm{n15}}$		    & 74.37   & 5,2 \\ 	
$i_{\mathrm{n12}}$                       & 257.21 & 2         &   $\mathrm{solar}+i_{\mathrm{n9}}$	    & 92.89   &  1,2\\   
$i_{\mathrm{n13}}$                       & 304.40 & 2         &   $\mathrm{solar}+i_{\mathrm{n10}}$	    & 94.82   &  1,2\\   
$i_{\mathrm{n14}}$                       & 463.06 & 2         &   $\mathrm{solar}+i_{\mathrm{n11}}$	    & 101.47  &  1,2\\   
$i_{\mathrm{n15}}$                       & 539.71 & 2         &   $\mathrm{solar}+i_{\mathrm{n12}}$	    & 103.30  &  1,2\\   
$s_{m2}$                                 & 85.90  & 3         &   $\mathrm{solar}+i_{\mathrm{n13}}$	    & 109.14  &  1,2\\   
$s_{m3}$                                 & 63.56  & 4         &   $\mathrm{solar}+i_{\mathrm{n14}}$	    & 101.85  &  1,2\\   
$s_{m5}$                                 & 274.66 & 5         &	  $\mathrm{solar}+i_{\mathrm{n15}}$	    & 94.44   &  1,2\\   
$s_{m3}+i_{\mathrm{n9}}$                 & 63.15  & 4,2	      &   $s_{m3}+r+i_{\mathrm{n9}}$		    & 63.00   & 4,6,2\\  
$s_{m3}+i_{\mathrm{n10}}$                & 62.94  & 4,2	      &   $s_{m3}+r+i_{\mathrm{n10}}$		    & 62.80   & 4,6,2\\  
$s_{m3}+i_{\mathrm{n11}}$                & 62.77  & 4,2	      &   $s_{m3}+r+i_{\mathrm{n11}}$		    & 62.72   & 4,6,2\\  
$s_{m3}+i_{\mathrm{n12}}$                & 61.23  & 4,2	      &   $s_{m3}+r+i_{\mathrm{n12}}$		    & 61.23   & 4,6,2\\  
$s_{m3}+i_{\mathrm{n13}}$                & 59.95  & 4,2	      &   $s_{m3}+r+i_{\mathrm{n13}}$		    & 59.92   & 4,6,2\\  
$s_{m3}+i_{\mathrm{n14}}$                & 59.71  & 4,2	      &   $s_{m3}+r+i_{\mathrm{n14}}$		    & 59.71   & 4,6,2\\  
$s_{m3}+i_{\mathrm{n15}}$                & 54.97  & 4,2	      &   $s_{m3}+r+i_{\mathrm{n15}}$		    & 54.57   & 4,6,2\\  
$s_{m3}+r$                               & 63.26  & 4,6	      &   2-step $i$			    &	  50.96    & 7 \\	
\hline
\hline
\end{tabular}
\tablefoot{References and model details:
(1): Solar abundances from \citet{Lodders2003}; 
(2) $i$-process abundances from \citet{Hampel2016}, using constant temperatures of $T=0.15\, \mathrm{GK}$ and constant densities of $\rho = 1600 \, \rm g \,cm^{-3}$. The  respective neutron densities are indicated (as $\log$N [cm$^{-3}$]) by the subscript; 
(3): AGB yields for M$_{\rm init}=2$ M$_{\odot}$, ${Z}=0.0001$ \citep{Lugaro2012}; (4): AGB yields for M$_{\rm init}=3$ M$_{\odot}$, ${Z}=0.0001$ \citep{Lugaro2012}; (5): AGB yields for M$_{\rm init}=5$ M$_{\odot}$, ${Z}=0.0001$ \citep{Lugaro2012};
(6): $r$-process from dynamical ejecta of binary neutron star merger \citep{Korobkin2012}; (7): $i$-process with two ingestion episodes
of $\tau$=0.30 and 0.96 mbarn$^{-1}$ (Sect.~5).} 
\end{table}
\subsection{$s$- versus $i$-process}
\begin{figure*}[htb]
\begin{center}
\includegraphics[angle=0,width=0.75\hsize]{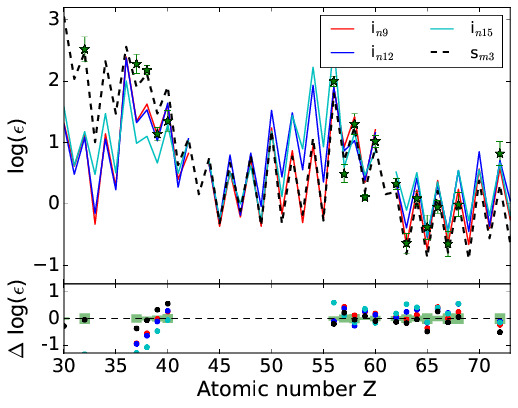}
\end{center}
\caption{Heavy element pattern in \#10464 in comparison with $i$-process model calculations for various neutron densities reaching from 
$N=10^9 \, \mathrm{cm}^{-3}$ to $N=10^{15} \, \mathrm{cm}^{-3}$. 
{Here, the $s$-process curve is for the best-fit 3 M$_{\odot}$ AGB 
composition of \citet{Lugaro2012}  (case ``$s_{m3}$''; Table~3).}
The top panel shows absolute abundances, the lower panel the resulting residuals of the individual fits, and the errorbars of the abundances as squares.
{ Pb (Z=82) was excluded from our statistical tests and is thus not shown in this and the following figures.}} 
\label{fig:individual_pattern}
\end{figure*}

Fig.~\ref{fig:individual_pattern} shows the {best-fit $s$-process results from the AGB models of \citet{Lugaro2012}, and also
different undiluted $i$-process models
with neutron densities of $N=10^{9}$ cm$^{-3}$ up to $10^{15}$ cm$^{-3}$} \citep{Hampel2016}. We 
did not attempt to fit a pure $r$-process pattern to the star given its higher metallicity where Galactic chemical evolution 
dictates that already several other sites have contributed
to its chemical enrichment.

Our least-squares fitting emphasizes that the pure, diluted $s$-process pattern of the \citet{Lugaro2012} 
yields already provide a good agreement with the observed data. 
Here, we find a progenitor with an initial mass (M$_{\rm init}$) of 
3.0 M$_{\odot}$ (M$_{\rm evol}$=2.51 M$_{\odot}$ after evolution including 20 thermal pulses) 
to provide the best fit of the observations\footnote{ 
Fitting the entire suite of F.R.U.I.T.Y. models yielded a lower AGB mass of 1.5 M$_{\odot}$, albeit at a poorer match in metallicity so that 
we did not pursue this comparison any further.};  
specifically it is characterized by core and envelope masses of \mbox{M$_{\rm Core}$  = 0.81 M$_{\odot}$} and
\mbox{M$_{\rm Env.}$ =  1.70 M$_{\odot}$}, respectively, also implying a fairly massive white dwarf companion. 

\begin{figure*}[htb]
\begin{center}
\includegraphics[angle=0,width=0.85\hsize]{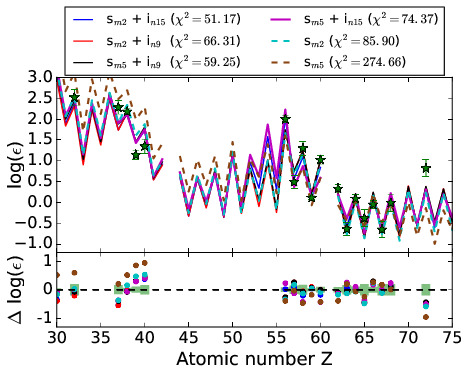}
\end{center}
\caption{Same as Fig.~\ref{fig:individual_pattern}, {but for various mixtures of $s$-process with $i$-process models. 
Shown are the curves for the overall, best-fit version ($s_{m2} + s_{n15}$; blue), a 2 M$_{\odot}$ AGB component plus a lower neutron density (red), 
and an $s$-process from a high-mass AGB model plus the two extreme neutron density $i$-processes (black and magenta). 
Shown as dashed lines are two $s$-process models from different AGB masses.}} 
\label{fig:mixed_with_agbi}
\end{figure*}

In comparison, the sole assumption of an $i$-process enrichment provides a larger $\chi^2$, which was smallest for 
a neutron density of $N=10^{9}$ cm$^{-3}$. 
The differences between model  and observations are minimal for the  second $s$-process peaks, while, for the light $s$-elements,  
this scenario only succeeds in reproducing either Y or Zr. 
{ Overall, the $\chi^2$ values in the hundreds indicate that these fits can be considered bad and statistically insignificant.
Here, it is worth pointing out that an overproduction of Pb, as described in Sect.~3.4, is also seen in the $s$-process calculations (e.g, Fig.~1) and therefore
not only inherent in the $i$-process models -- conversely, the latter can actually aid to help with solving these problems in a bigger frame, although
this endeavour is far beyond the scope of this work in a single, peculiar star.}

\citet{Denissenkov2017} suggested that the $i$-process in rapidly accreting white dwarfs can account for about a third of the  intermediate $n$-capture elements
(32$\le$Z$\le$42) within Galactic chemical evolution. 
The abundance pattern from their model {does not match} our observations in the bulge star \#10464 as none of the element abundances
seen in this star reaches the over-enhancements predicted in that scenario (cf. their Fig.~4). We note, however, that the
respective calculations have been carried out for explicitly higher metallicities ([Fe/H]$> -1$ dex) than the one found in this bulge object.

Based exclusively on the high Rb/Zr ratio, \citet{Koch2016} suggested that the AGB progenitor was likely of intermediate mass, at $\sim$4 M$_{\odot}$, although
a detailed match of the remaining abundance pattern (of 10 elements with Z$\ge$30) could not be reached. 
Similarly, either set of models employed in the present work 
fails to make sense of the very high [Rb/Fe] abundance of this star \citep[cf.][]{Abia2001}. 
Even more severe is the trend of strongly decreasing [$ls$/Fe] ratios when moving from Rb to Zr (Fig.~1). This is not reproduced in any of our simulations
and poses a challenge to nucleosynthetic calculations. 
A decrease from Sr through Zr is seen in models of fast rotating massive ($\sim$25 M$_{\odot}$) stars \citep{Frischknecht2012}, yet those have the tendency to 
produce low amounts of Rb. 
The latter is predominantly produced in AGB stars, but constructing a superposition of such enrichment with the more regular 
intermediate-mass pollution \citep{vanRaai2012} to account for \#10464's heavy element patterns seems unlikely. 
High Rb can also be indicative of high neutron-density \citep{Pignatari2010}. 
In turn, comparison with the models of \citet{PerezMesa2017} indicate that $\sim$6 M$_{\odot}$ AGB star can indeed 
produce the high, observed Rb abundance, but this conflicts with a too high Rb/Zr ratio of our observations.
{ Indeed, our fitting of a pure higher-mass (5 M$_{\odot}$) AGB $s-$process component (``$s_{m5}$'' in Fig.~3 and Table~3)
leads to a deterioration of our statistics.}

If we only assume a short neutron burst in our $i$-process calculations, which does not provide enough neutrons to establish a typical equilibrium-abundance pattern, the heavy element production is only driven up to the $ls$ peak. In such a scenario, a neutron density of $N=10^{9}$ cm$^{-3}$, leading to an exposure of 
$\tau=0.3$ mbarn$^{-1}$, can reproduce the observed characteristics of the Rb peak. 
However, any further neutron irradiation, as is needed for the production of heavier elements including the $hs$-peak elements and Pb, would destroy the reproduced ls pattern. Therefore it cannot be assumed that the $i$-process can produce both the Rb peak and the heavier elements in one single event. 

As for the second-peak elements, none of the models we employed is able to reproduce the shape of the heavy-$s$ peak (e.g., the observed [Ba/La] ratio), which 
renders a pure $s$-process origin unlikely; this is, e.g., manifested in the [Ba/La] vs. [Eu/La] plane \citep[Fig.~6 in][]{Mishenina2015}, where 
our star grazes the lowest boundary of open cluster and Galactic disk stars' [Eu/La] values.
While models of the $i$-process generally succeed in reproducing a higher [Ba/La] compared to the $s$- or $r$-process \citet[e.g][]{Hampel2016}, the observed Ba/La ratio of \#10464 is, per se, too low for 
a substantial $i$-process contribution characterised by $N=10^{15}$ cm$^{-3}$. Moreover, the shape of the $hs$-peak is remarkable, since [Ba/La]$\gg$[Ba/Ce]. 
An increased neutron density shapes the $hs$-peak predominantly through contributions of additional Ba resulting from the decay of radioactive $^{135}$I, 
which, however,  has trouble explaining both the Ba- and Ce-to-La ratio being 0.3 dex higher than solar.
\section{Multiple enrichment sites}
As we have shown in the previous section, it is hard to reconcile the observed heavy-element abundance pattern in 
\#10464 with only one nucleosynthetic event at a time. 
Therefore, {in the following we will focus} on exploring the possibility that this bulge object was enriched by more than one progenitor, each 
having contributed some fraction of {the} two or three nucleosynthesis processes  described above.
In order to  fit these processes to the observed abundance distribution of the star, we adopt a linear superposition of $N$ individual nucleosynthetic 
processes, $j$, following the  formalism of \citet{CJHansen2014}:
\begin{equation}
Y_{\rm calc}(Z) =  \sum_{j=1}^N c_j  Y_j(Z), \qquad \text{with } c_j \ge 0
\end{equation}
where $Y_j(Z)$ denotes the absolute abundances and $c_j$ are the weights assigned to each of the  contributing processes ($s$,$r$,$i$), respectively. 
Since $Y_j$ includes an arbitrary scaling factor, the actual values of these weights have no physical meaning. {We note that implementing $N$ weights without additional constraints will include an additional additive freedom. However, there are other mixing techniques as shown in, e.g., \citet{Hampel2016}, where the weighting factors are constrained by $\sum c_j = 1$. We want to stress that the choice of the mixing techniques does not affect the conclusion of this work.}
In the minimization process, $Y$ was substituted by $Y_{\rm calc}$ in eq.~1. 
If more than three different processes were included, convergence could not be achieved.  
The results for $Z_i$ = 32 and $Z_f$ = 72 are again indicated in Table~3.

{In the following, we used the entire set of  $s$-process patterns from \citet{Lugaro2012} as described above (Sects.~3.1, 4.1), 
the Solar abundance scale from \citet{Lodders2003}, the diluted $i$-process pattern from \citet{Hampel2016} (Sect. 3.4), 
and the theoretical $r$-process, calculated from the dynamical ejecta of a neutron star merger (Sect.~3.2). 
In total, more than 10000 different model combinations were thus tested. 
In Figs.~3 and 4 we show the observed abundances $\log \epsilon$ for  the star \#10464 together with several exemplary combinations, and the best-fit 
linear combinations of the various processes.}
\begin{figure}
\begin{center}
\includegraphics[angle=0,width=1\hsize]{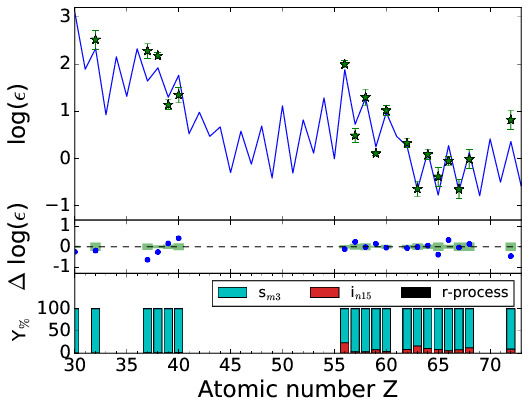}
\end{center}
\caption{Same as Fig.~2, but for a linear combination of all three nucleosynthetic channels ($s+r+i$).
The bottom panel indicates the relative contributions from each process. Note that no $r$-process component is required.}
\label{fig:three_mixture}
\end{figure}

As a result,  the linear admixture of other processes to the $s$-process prescription improve the fits  slightly.
If we {\em a priori} assume that the $s$-process must come from the same, fiducial source as derived above from a single site
(viz., a metal-poor 3 M$_{\odot}$ AGB star) and the $i$-process would act as a mere perturbation on top of the AGB yields, 
we need to invoke the highest tested neutron density for a ``best'' match (case $s_{m3}+i_{n15}$ in Table~3). 
However, an even better $\chi^2$ was obtained for the case of an $s$-process from a 2 M$_{\odot}$ AGB star plus 
the highest $n$-density ($i_{n15}$) $i$-process.

Conversely,  the most neutron rich scenario we tested, the $r$-process, leads again to no significant improvement of the $\chi^2$ (labeled ``$s+r+i$''). 
Moreover, for three production processes, the best fit was achieved without any fraction of the $r$-process, thereby 
leading us back to the above $s+i$ scenarios. 
This can be seen in the lower panel of Fig.~4, where the contributions of the specific processes to each isotope are shown. 
The $r$-process does not contribute to any isotope, but a small fraction of $i$-process is seen to contribute 
to the region between $56\le Z \le 72$. Note that, still,  none of the mixtures are able to reproduce the high amount of Rb and Sr.

In the framework of considering multiple individual enrichment events, it is {standing to reason to consider the occurrence} of two distinct proton-ingestion events
in the same donor, each of different strength. 
Our current understanding of the {site(s)} of the $i$-process does not allow us to make firm constraints on the exact number of successive proton-ingestion events and it has been shown that, 
for example, super-AGB stars could host multiple such events \citep{Jones2016iproc}. 
While a shorter neutron bursts with $\tau=0.30$ mbarn$^{-1}$ can reproduce the light-$s$ peak,  adding a separate event with $\tau=0.96$ mbarn$^{-1}$ gives the best fit to the 
observed abundances of elements with $Z>50$. 
 Two separate bursts are required because the peak abundance moves to higher 
Z as the exposure increases, building up first the light s peak but then moving on to the
heavy s peak. If the exposure is high enough, Pb is built up. This is similar to the way heavy 
elements are produced in the $s$-process.

A combination of these two individual events thereby leads to the overall,  best { (in a $\chi^2$-sense)} explanation of the peculiar abundance 
pattern of \#10464  (Fig.~5), 
although the high complexity of this scenario renders it, statistically, equally (im-)probable as a 3\,M$_{\odot}$ AGB pollution plus single $i$-process 
contributions. At respective $\chi^2$ values on the order of 50 vs. 60, the differences are marginal.
Typically, in abundance fitting excellent $\chi^2$ statistics as low as $\sim$10 can be reached \citep[see also][]{CJHansen2014,Abate2015stat}. 
Our higher values in Table~3 indicate that the composition of this star is not fully understood, yet, 
{ and cannot be explained satisfactorily with 
any of the processes considered, or combinations thereof.}
\begin{figure}
\begin{center}
\includegraphics[angle=0,width=1\hsize]{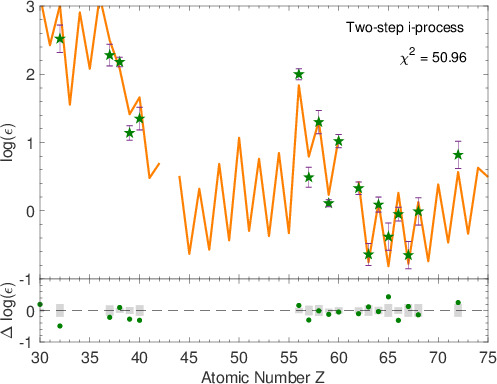}
\end{center}
\caption{Same as Fig.~2, but for a two-step $i$-process with two separate ingestion events of $\tau=0.30$ and 0.96 mbarn$^{-1}$.
Such a superposition is better able to reproduce both the light and heavy $s$-peaks.}
\end{figure}
\section{The impact of self-pollution}
The stellar parameters of T$_{\rm eff}$=5400 K and log\,$g$=1.7 derived by \citet{Koch2016} place this star on the 
horizontal branch, where evolutionary tracks indicate a mass of $\sim$0.55 M$_{\odot}$ \citep[e.g.,][]{Cassisi2004,CJHansen2011}. 
As the spectroscopic gravities of the sample of \citet{Koch2016} were based on accurate ionization equilibrium, also the stars' distances
could be determined; in turn, we estimate that star \#10464 has a luminosity of  $\sim$220~L$_{\odot}$. 

At this evolved level it is likely that this star has undergone deep evolutionary mixing toward the tip of the RGB, which 
will have altered its surface composition. For the case of carbon, this can be quantified using the evolutionary 
calculations of \citet{Placco2014}, which suggest a upward correction in [C/Fe] on the order of 0.2 dex, bringing the carbon ratio
of \#10464 to $\sim$0.6 dex. These effects were also recently discussed by \citet{Henkel2018} in the context of an improved formalism for thermohaline mixing 
in metal-poor stars.

In addition to the moderately enhanced carbon-level of this star, we found a strong enhancement in nitrogen (Table~1), resulting in 
a [C/N] ratio of  $-0.34\pm0.26$ dex, or, accounting for the aforementioned correction for stellar evolution, [C/N]$\sim -0.54$ dex. 
This value is close to the boundary of $-0.6$ dex that separates evolved, mixed stars from objects that are unaffected by mixing \citep{Spite2005,CJHansen2016}. 

\begin{figure}
\begin{center}
\includegraphics[angle=0,width=1\hsize]{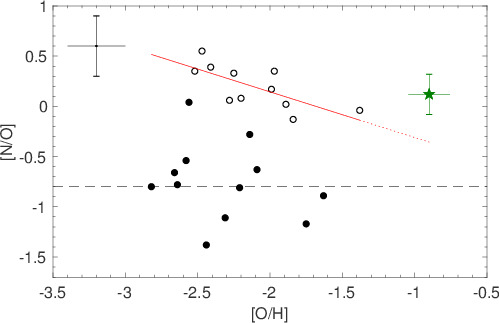}
\end{center}
\caption{N and O abundances in metal-poor halo stars from \citet{Spite2005}. A typical error bar is indicated top left; the bulge star 10464 is shown as 
a green star symbol. Solid (open) symbols indicate unmixed (mixed) stars and the dashed line illustrates the mean value of the unmixed sample.
The red line is the best-fit relation for the mixed stars, determined by \citet{Spite2005}, extrapolated toward our bulge star (dotted line).}
\end{figure}
This is strongly manifested in Fig.~6, where we distinguish mixed and unmixed stars in the metal-poor halo sample of \citet[][their Fig.~11]{Spite2005}.
Here, the fact that mixed stars  have  converted C to N in the CN cycle is seen through their 
 systematically higher [N/O]. 
 \citet{Spite2005} also note a strong correlation between the [N/O] and [O/H] ratios in the mixed stars (solid/dotted lines in Fig.~6), which is not seen in the unmixed counterparts 
 (dashed line). 
This is due to an overabundance of N in the surface of the mixed stars on top of the large, initial abundance spread. Assuming that this extra, secondary nitrogen is independent of the stellar metallicity, the [N/O] ratio would decrease with increasing metallicity, [O/H], as is seen in the observed abundances for the mixed stars. 
The bulge star 10464 lies marginally above the extrapolated trend of the mixed halo stars. Nonetheless, its elevated [N/O] ratio places it uniquely in the regime of 
mixing.

A strong level of self-pollution would indicate that the presently observed surface abundance has been significantly altered from its initial composition.
Thus it is possible that the abundance peculiarities seen in \#10464 do not reflect an external polluter's signatures only, aggravating a meaningful 
comparison with the models as described in the previous sections.
The extent to which the signatures of an external polluter are disguised by mixing events depends on how the pollution has occurred. If the pollution was already present in the gas from which the star formed, then evolutionary mixing only plays a minor role in altering the composition. While on the main sequence, settling and levitation may change the surface layers, but these effects are removed once a deep convective envelope starts to develop 
\citep{Richard2002,Matrozis2016}.
Processing of material near the tip of the giant branch only affects the lighter elements \citep{Gratton2000,Stancliffe2009}, 
with oxygen and beyond remaining unaffected. The heavy elements should therefore be representative of the material the gas formed from.

When pollution occurs from a companion star, the situation is more complex. Accreted material, which has undergone nuclear processing in the companion, will have a higher mean molecular weight than the unprocessed material of the star that receives it. The accreted layer will thus be unstable to thermohaline mixing, which has the effect of mixing the accreted layer into the recipient's interior very rapidly, typically a small fraction of the main-sequence lifetime \citep{Stancliffe2007}. In addition, rotation may also cause mixing of accreted material \citep{Matrozis2017}. If the accreted material is mixed to less than the depth that the convective envelope reaches during the ascent of the main sequence (roughly 0.45 M$_\odot$; \citealt{Stancliffe2008})
 further dilution will occur. After this, the light element surface abundances can still be changed by processing of material near the tip of the giant branch, as described in the previous paragraph, while the heavy elements will all have been diluted to the same extent. 
\section{Summary and conclusions}
An investigation of several nucleosynthetic models indicated that the abundance distribution of the peculiar bulge CH-star \#10464 
cannot be satisfactorily explained by pure AGB $s$-process nucleosynthesis, nor with a single $i$-process 
under conditions as derived from an AGB star in 
\citet{Hampel2016}. 
Its abundances are better fit by combinations of several nucleosynthetic  processes. 

Our study suggested that, if the enhancements were due to $s$-process pollution, 
  the progenitor of  this component  was likely a low-to-intermediate mass
AGB, in line with our conclusions in \citet{Koch2016} from the F.R.U.I.T.Y. database \citep{Cristallo2011}, although
those findings were restricted to an assessment of the Rb/Zr and $hs$/$ls$ ratios. 
Similarly, the aforementioned CEMP-$s$ star 27793 was found by \citet{Koch2016} to have been enriched by a $\sim$4 M$_{\odot}$ AGB star, 
although the predictions of \citet{Abate2015models} suggest lower-mass companions for CEMP-$s$ stars of around  0.9--1.1 M$_{\odot}$. 
The fact that our abundance matching results in  fairly common progenitor masses implies that AGB  companions to such metal-poor bulge stars
were very similar in nature to the present-day, metal-rich bulge AGB population \citep[e.g.,][]{Uttenthaler2015}. 

 It is more likely (from a mere statistical point of view) that the abundance pattern in this star was  caused by some $i$-process nucleosynthesis, albeit a more complex scenario 
 than the simple picture including one ingestion event (Sect.~3.3.) 
 cannot statistically be ruled out.
In a single event, mass conservation would dictate a decline in the second-peak elements (around Ba) accompanying
an enhancement in the first peak elements (such as Rb and Sr), and vice versa. 
This is in contrast to the high, relative strength of the light and heavy neutron-capture peaks, indicating the occurrence of at least two
ingestion periods.

A zoo of other processes to have entertained the enrichment of this star is certainly conceivable, such as electron-capture SNe 
(at a similar outcome as the weak-$r$ process), $\nu$-driven winds, $\nu p$-processes, or an $\alpha$-rich freeze-out.
However, a decomposition of the heavy element pattern into two components -- an AGB-dominated $s$-process with an 
admixture of  $r$-process rich ejecta from neutron star mergers --
already did not yield any significant improvement of the statistics. 

More data for this class of stars are clearly needed, but this request comes at a price:
most of the models considered here are most sensitive to heavy elements that are notoriously difficult to measure, such as Os or Ir, whereas most of the  dominant heavy element transitions lie  predominantly in the blue-to-UV spectral range \citep[see also][]{CJHansen2015}, which is challenging for anything but the metal-poor halo  
\citep[e.g.,][]{Roederer2016iproc}. 
\begin{acknowledgements}
We are very grateful to M. Pignatari, and also J. Bliss,  
and C. Ritter for helpful discussions. 
The anonymous referee is thanked for a fast and constructive report, and M. Hanke for
support with the ATHOS code.
This work was supported by Sonderforschungsbereich SFB 881 "The Milky Way System"  (subproject A04) of the German Research Foundation (DFG).
M.R. and A.A. acknowledge support by the Helmholtz-University Young Investigator grant No. VH-NG-825 and ERC Starting Grant 677912 EUROPIUM. 
R.J.S. acknowledges the support of STFC through the University of Hull Consolidated Grant ST/R000840/1.
\end{acknowledgements}
\bibliographystyle{aa} 
\bibliography{ms} 
\end{document}